\documentclass[journal]{IEEEtran}

\usepackage{cite}
\usepackage{amsmath,amssymb,amsfonts}
\usepackage{algorithmic}
\usepackage{graphicx,color}
\usepackage{textcomp}
\usepackage{xcolor}
\usepackage{graphicx}   
\usepackage{subcaption} 
\usepackage{float}
\usepackage{microtype} 
\usepackage{booktabs} 

\usepackage{hyperref}

\usepackage{amsmath}
\usepackage{amssymb}
\usepackage{mathtools}
\usepackage{amsthm}

\usepackage[capitalize,noabbrev]{cleveref}

\theoremstyle{plain}
\newtheorem{theorem}{Theorem}[section]

\theoremstyle{definition}

\theoremstyle{remark}
\newtheorem{remark}[theorem]{Remark}

\usepackage[textsize=tiny]{todonotes}

\begin{document}

\title{Block Erasure-Aware Semantic Multimedia Compression via JSCC Autoencoder}

\author{Homa Esfahanizadeh, \textit{Member, IEEE}, 
Nargis Fayaz, \textit{Graduate Student Member, IEEE},
Jinfeng Du, \textit{Senior Member, IEEE}, and Harish Viswanathan, \textit{Fellow, IEEE}\\
\thanks{Homa Esfahanizadeh, Jinfeng Du, and Harish Viswanathan are with Nokia Bell Labs, Murray Hill, NJ 07974, USA. Nargis Fayaz is with Nokia Bell Labs, Bangalore, India.
E-mails: \{homa.esfahanizadeh, jinfeng.du, harish.viswanathan\}@nokia-bell-labs.com, eez218533@ee.iitd.ac.in }}

\maketitle


\begin{abstract}
We present an AI-based framework for semantic transmission of multimedia data over band-limited, time-varying channels. The method targets scenarios where large content is split into multiple packets, with an unknown number potentially dropped due to channel impairments. Using joint source–channel coding (JSCC), our approach achieves reliable semantic reconstruction with graceful quality degradation as channel conditions worsen, eliminating the need for retransmissions that cause unacceptable delays in latency-sensitive applications such as video conferencing and robotic control. The framework is compatible with existing network protocols and further enables intelligent congestion control and unequal error protection. A tunable design parameter allows balancing robustness at low channel quality against fidelity at high channel quality. Experiments demonstrate significant robustness improvement over state-of-the-art baselines in both image and video domains.
\end{abstract}


\section{Introduction}

{Semantic transmission of multimedia over wireless networks aims to deliver critical information (i.e., its semantics) with minimal latency while utilizing context at the receiver to ensure task effectiveness. Traditional transmission protocols over network like TCP rely on retransmissions when packets fail, introducing delays that are unsuitable for time-sensitive applications such as video conferencing or remote robotic control. In these scenarios, timely reconstruction with slightly lower quality is often preferred over delayed, higher-quality transmission. Furthermore, conventional compression methods like H.264/H.265 or JPEG-based codecs employ entropy coding for compression and assume lossless packet transmission, leading to significantly degraded multimedia content when retransmissions are not possible. This challenge is further intensified in time-variant and heterogeneous networks, where stringent latency requirements prevent adjusting the code for every channel snapshot or user.}

\begin{figure}
    \centering
    \includegraphics[width=0.99\linewidth]{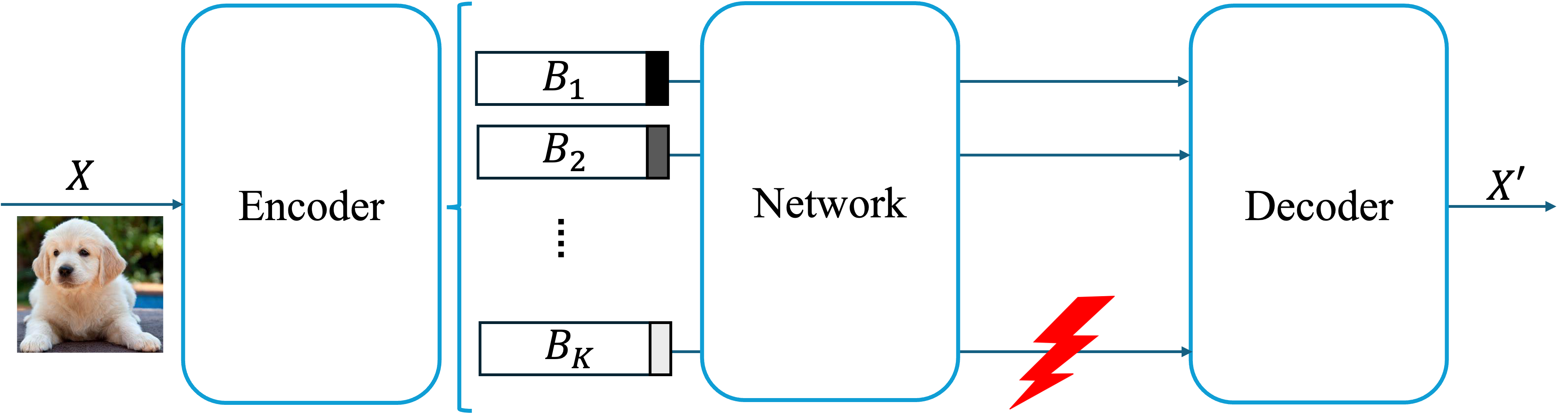}
    \caption{Our encoder converts the multimedia sample into  blocks with predefined importance levels. Here, darker colors indicate blocks that are more important for reconstruction. Depending on the bandwidth and channel quality, the network delivers a subset of blocks to the decoder, where the original sample is approximately reconstructed from the received subset. \vspace{-0.5cm} }
    \label{fig:methodology}
\end{figure}

In this direction, we propose a new AI-enabled pair of source encoder and decoder, based on joint source-channel coding (JSCC). This solution allows the network to drop data blocks under bandwidth constraints while still allowing the decoder to reconstruct a lossy version of the original sample from a subset of blocks. The solution differs from traditional hierarchical coding or layered encoding at the source and successive refinement or error concealment at the destination, e.g., \cite{Cover1991, Rose2003, SuccessiveRefinementImageDeep, Harish2000, Foad2022}, by explicitly taking the stochastic loss into account during the design of the encoder/decoder through joint source-channel coding. The overall methodology is illustrated in Fig.~\ref{fig:methodology}. 
This innovation enables the network to handle data semantically under harsh conditions through intelligent congestion control, unequal error protection (UEP) at the code-block level, or by avoiding retransmissions. In particular, UEP allows JSCC to be applied at two stages, i.e., at the source encoder as well as at the channel encoder, since the importance levels convey information about the source to the channel, enabling different treatment across levels. Although not strictly joint coding, this results in source distribution–specific behavior at the channel, while remaining compatible with current networks.
 

Our JSCC code is built on a block-erasure–based architecture that follows a novel guideline for handling different parts of an encoded message at the granularity of blocks rather than individual bits. In contrast to prior work \cite{TungTCOM2024}, and closer to real network behavior, we model channel impairment as block erasures instead of bit errors. This design avoids forcing the system to process erroneous bit sequences to achieve semantic gains. To support this setting, we propose two new training techniques: (i) stochastic modeling of block-erasure events, and (ii) applying the quantizer only during inference. Together, these enable JSCC to operate effectively over multi-level block-erasure channels. A key advantage of our approach is that it requires only a single encoder–decoder pair that adapts to varying distortion levels, whereas prior work often depends on multiple encoder–decoder pairs \cite{Foad2022,khosravirad2025ratelessjointsourcechannelcoding}.
\section{Related work}

We categorize the relevant state-of-the-art into four groups, explain each category, and highlight how our proposed method differs from them.

\subsection{Joint source channel coding (JSCC)}

Joint source and channel coding (JSCC) is the process of jointly performing compression (i.e., source coding) and error correction (i.e., channel coding). Source coding reduces the intrinsic redundancies within a sample for compression, and channel coding adds redundancies for error correction. JSCC, by combining these two operations, is known to outperform separate source and channel coding designs in finite block lengths \cite{kostina13_lossy_joint_sourc_chann_codin}. However, its practical adoption has remained limited due to the complexity of its problem formulation and implementation requirements. Recent advances in deep learning have led to the development of superior JSCC methods achieving larger benefits, e.g., \cite{bourtsoulatze19_deep_joint_sourc_chann_codin,tung22_deepw,xie21_deep_learn_enabl_seman_commun_system} among others. Nevertheless, despite their design improvements, these methods face several challenges when applied in real-world networks: (i) the channel distribution is typically unavailable at the source, making it difficult to design an effective JSCC scheme; (ii) they eliminate the role of channel coding and modulation, which typically operate independently of the source at the transmitter; (iii) they assume erroneous packets can be received by upper layers at the receiver, contradicting the layered architecture of current OSI stack model.

A recent method attempts to address the first two challenges stated above. They do so by designing the JSCC code at the source for a specific abstraction of the channel  as a multi-level bit error rate (BER) interface, while delegating the task of shaping the actual loss distribution to the one described by the interface to the network—bringing it closer to the conditions under which the code was trained \cite{TungTCOM2024}. {Its performance can be further improved by making the interface parameters trainable  \cite{learned-MLRI}.}  However, this approach still struggles with the third challenge: operating in the bit domain often requires the acceptance of erroneous packets, which violates OSI protocol stack rules. In our proposed work, the network is not required to pass erroneous packets on to the image or video decoder. Instead, errors are marked as block erasure(s) and unequal protection at the block level is delegated to the network in a manner that aligns more closely with existing networking principles. In essence, the multi-level block erasure channel abstracts the actual channel, enabling joint source channel coding at the source. The transmitter at the bottleneck link receives packets with encoded and possibly encrypted data corresponding to different importance levels, thereby abstracting the source distribution to multiple-levels of uniform distribution from the transmitter's perspective. This abstraction allows the transmitter to also employ ``joint" treatment of  source and channel through UEP or  intelligent packet dropping. Consequently, JSCC becomes practical through a JSCC at the source encoder for an abstracted channel and a JSCC at the channel encoder for an abstracted source.

Finally, there exists a line of work that trains autoencoders for video transmission equipped with packet-level error concealment, such as GRACE \cite{GRACE}. The proposed approach differs from GRACE in two key aspects. First, GRACE spreads the impact of dropped packets via pseudo-random masking to conceal these erased elements, whereas our method operates as a packet-level erasure code in which the decoder has explicit knowledge of the erasure locations. Second, in GRACE all encoded packets have the same importance, whereas in our model the importance follows a predefined interface between the application and the network, facilitating differentiated quality of service provisioning by the network.

\subsection{Successive refinement}

Successive refinement (hierarchical/progressive compression) is used to compress the original sample into a base layer and several ordered refinement layers \cite{Cover1991, Rose2003, SuccessiveRefinementImageDeep, Harish2000, Foad2022,khosravirad2025ratelessjointsourcechannelcoding}. Depending on network conditions, more bit budget or bandwidth is allocated to the base and lower-order refinement layers compared to higher-order ones. However, conventional methods apply greedy coding—where the base layer represents the best possible approximation under a most limited bit budget, and each subsequent layer improves upon the previous representation. Our method, instead, enables a more flexible trade-off: even the most critical part of the representation accounts for the possibility of future layers contributing collaboratively to a better overall reconstruction. As a result, our approach achieves improved performance and offers a tunable trade-off between resolution under harsh channel conditions and optimal quality in good channel conditions. Moreover, those codecs are often exclusively compression-based, without adding redundancy to any layer. In contrast, our approach incorporates redundancy into each layer, as packet losses in the network are considered at the time of compression through JSCC for an abstracted channel. Lastly, this line of work typically requires designing multiple decoders, one per resolution, which induces a large complexity and overhead when the number of resolution layers is high.

\subsection{In-painting}
Inpainting, a task in computer vision, involves filling in masked regions of an image with new content, e.g., \cite{9878449,9157623}. While this technique has been primarily explored for images, it can potentially be extended to other types of source distributions. Conventionally, inpainting methods do not introduce additional redundancy to compensate for block erasures; instead, they rely on the intrinsic redundancies within the image. As such, they do not fall under the category of JSCC. Moreover, these methods often produce high-quality reconstructions, but at the cost of possible hallucination due to the use of generative models—an outcome that contradicts the principles of communication systems. In contrast, our JSCC-based approach avoids hallucination and instead produces lower-fidelity reconstructions of the original sample, staying faithful to the received data, desired by a communication system.

\subsection{Rateless compression using dropouts}

Deep learning (DL)-based compression with variable rates using autoencoders has been studied \cite{Ng_NA_13976,9174523}, where the latent space is subjected to random dropouts. In these methods, an autoencoder is trained with an expanded latent space. During training, a random subset of encoded nodes is masked (set to zero), teaching the decoder to reconstruct a lossy version of the original sample by replacing the erased bits with zeros. In \cite{Ng_NA_13976}, dropout is applied uniformly at random, while in \cite{9174523}, it occurs stochastically from the tail of the latent space. The key differences between our proposed method and the prior approaches are as follows: (i) While prior approaches also fall under joint source-channel coding (JSCC), our proposed method targets a different channel model than them—specifically, i.e., a block erasure channel—which allows us to address block/packet erasures occurring during communication. (ii) Prior methods are effective mainly for contiguous erasures from the tail of the latent space, whereas our approach considers a non-zero probability of erasure at any block position. (iii) In previous work, the location of erasures is not communicated to the decoder (the erased bits are simply zeroed during training, which is a valid value), whereas in our method, a designated value is used to represent erasures. This allows the decoder to explicitly identify and handle erasures, leading to more effective reconstruction.

{Similarly, the ResiTok framework \cite{ResiTok,ResiVidTok} uses zero-out training to simulate token loss, enabling progressive encoding by prioritizing semantically critical information in earlier tokens of an image or video frame. However, it relies on successive refinement, restricting erasures to the end of the token sequence.}

\section{Method}

We propose a block erasure-aware semantic multimedia encoding scheme in which the content is divided into semantic blocks — each potentially varying in importance — and intrinsic redundancy is embedded across blocks. This enables the decoder to reconstruct an approximate version of the original message using only a subset of received blocks. Distortion is minimized when the subset includes semantically important blocks. Our approach is based on JSCC over a set of block erasure channels, a design space that remains largely unexplored in existing literature. This framework allows the network to cooperate with the source to prioritize transmission of semantically critical information without knowing the semantics. Some of the network operations that can benefit from our proposed solution are:
\begin{itemize}
    \item Unequal error protection: Blocks with higher semantic importance are better protected, for example, using stronger modulation and coding schemes (MCS), while less important blocks receive weaker protection.
    \item Intelligent congestion control: When network experiences congestion and must drop packets, it prioritizes the retention of more important blocks.
    \item Selective retransmission: The receiver only requests retransmission of semantically critical packets, reducing latency and improving goodput at the cost of tolerating some distortion. Less important missing packets are simply treated as erasures during reconstruction.
\end{itemize}

In summary, our proposed method facilitates efficient transmission of delay-sensitive content over heterogeneous and time-varying networks by accepting controlled distortion and prioritizing semantically important content under bandwidth constraints. 

Our source code (pair of encoder and decoder) is designed following specific channel abstraction (called interface defined by block erasure rates between the encoder and the decoder) to semantically encode the data sample into several blocks of various importance, including redundancies within each block to offer resilience against errors. 
A key benefit is a tunable trade-off mechanism that balances robustness under poor channel conditions against optimal performance in favorable conditions. Our solution facilitates a communication system where the network treats blocks of each multimedia sample based on the pre-defined specific abstraction (interface). As a result, source and network can collaborate using the shared guidelines to deliver a good performance under degraded channel conditions. Next, we describe the new solution and its benefits:

\subsection{Code design using JSCC trained for block erasure channel}

The encoder, implemented as a JSCC encoder, transforms the input sample $X$ into $K$ blocks, including some redundancy within the blocks. For simplicity, we are assuming each block is sent over a single packet, but in general a block can be sent over multiple packets. Due to bandwidth constraints, some blocks may be dropped by the network. Each block is pre-assigned an importance level, indicating the extent to which the reconstruction loss would increase if that block was dropped and not received. The behavior of this loss and its implications are discussed further in Section~\ref{sec:applications}, where specific applications are explored. The decoder reconstructs the sample using the subset of blocks that are successfully received, see Fig.~\ref{fig:methodology}.

\begin{figure}
    \centering
    \includegraphics[width=0.99\linewidth]{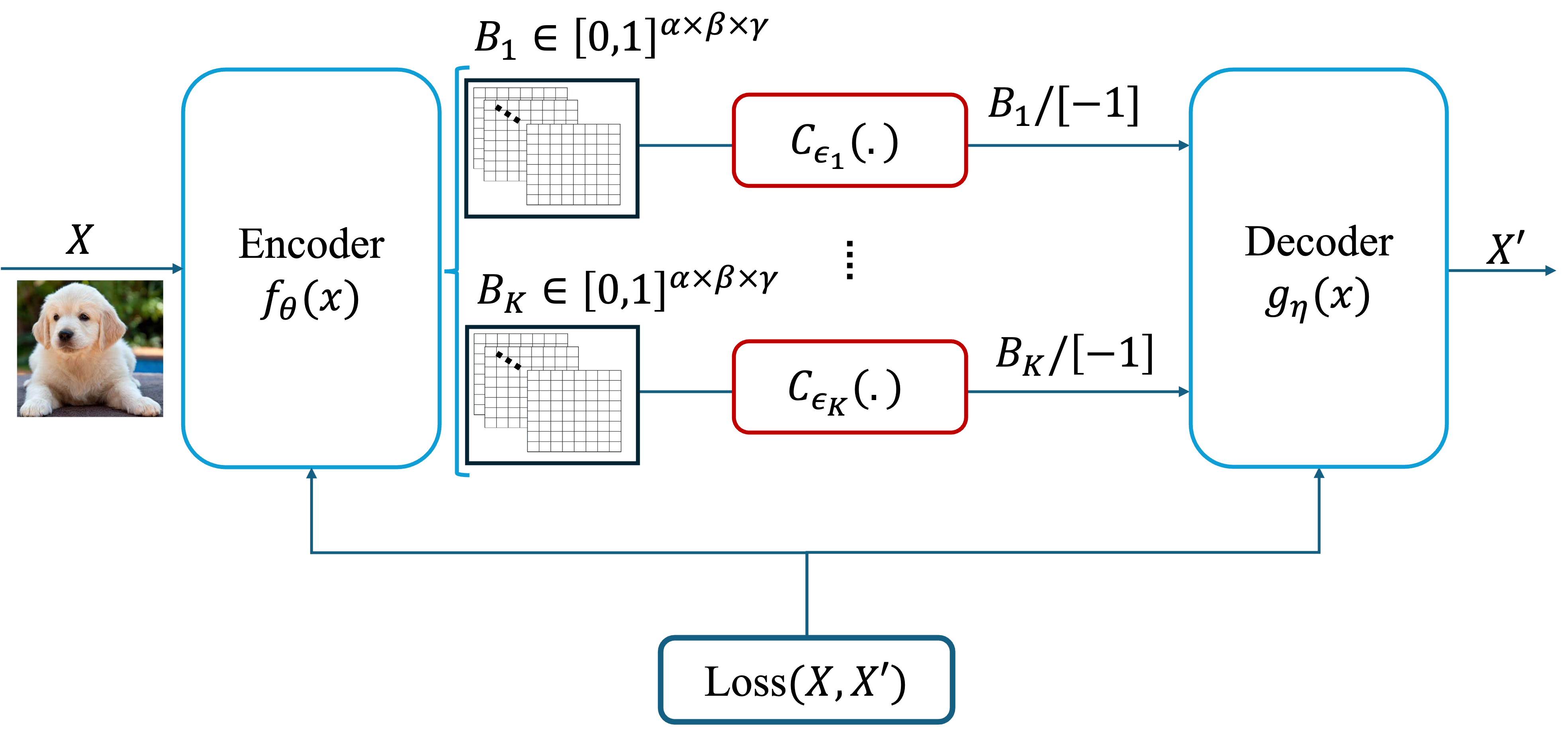}
    \caption{Designing the JSCC using a multi-level block erasure abstracted channel, enabling intelligent error concealment.}
    \label{fig:training}
\end{figure}

Next, we describe the training procedure for designing our encoder-decoder pair, as illustrated in Fig.~\ref{fig:training} in the context of images, though applies to other media such as video and audio. The neural encoder $f_\theta(X)$ transforms the input image $X$ into $K\alpha$ arrays of real numbers, each of size $\beta \times \gamma$, with entries in the interval $[0,1]$. These arrays are evenly grouped into $K$ blocks, so that each block $B_i$, $i \in \{1, \dots, K\}$, has dimensions $\alpha \times \beta \times \gamma$. Each block $B_i$ is then passed through an erasure channel $C_{\epsilon_i}(\cdot)$, which either preserves or erases the block entirely. In the event of erasure (with probability $\epsilon_i$), the decoder replaces the corresponding input with a placeholder block of the same size filled with the value $-1$:
\begin{equation}
    C_{\epsilon_i}(B_i)=\begin{cases}
        B_i,&\text{prob.}\;\; 1-\epsilon_i,\\
        [-1]_{\alpha\times\beta\times\gamma},&\text{prob.}\;\; \epsilon_i.
    \end{cases}
\end{equation}

\begin{remark}
    The use of $-1$, which lies outside the normal value range of $[0,1]$, enables the fixed-input-size decoder to identify which blocks have been erased.
\end{remark}

\begin{remark}
    To implement the erasure channels $C_{\epsilon_i}$ for $i \in \{1, \dots, K\}$ in a differentiable training setting, we adopt the following technique: for each $i$, draw a binary random variable $r_i$ such that $r_i = 0$ with probability $\epsilon_i$ and $r_i = 1$ otherwise. Then, $C_{\epsilon_i}(B_i)$ can be expressed as $r_i(B_i + 1) - 1$.
\end{remark}
\begin{remark}
    During training, we work with real numbers, where the elements of the latent space are real numbers within the interval $[0,1]$. However, during the inference (when utilized), each element is represented with $8$ bits, which has been shown (through extensive empirical evaluations) to be sufficient to preserve the performance of the encoder-decoder.
\end{remark}

Finally, the decoder reconstructs the sample from the received data, which may include erased blocks. The weights of the neural encoder and decoder are optimized during training to minimize the reconstruction loss between the original sample $X$ and the reconstructed sample $X'$.

\subsection{Tunable trade-offs across network operating regimes}

The vector $\epsilon$  specifies the erasure probability for each block of the encoded sample during training. Informally, this vector guides the source on how much information to allocate to each block, thereby shaping how sensitive the downstream task is to the erasure of individual blocks. By carefully designing this vector, the system can be optimized for operation under poor channel conditions, favorable conditions, or a continuum in between. We consider two representative scenarios:
\begin{enumerate}
    \item When $\epsilon_1=\dots=\epsilon_K=\epsilon$: In this case, a larger $\epsilon$ makes the source code to maintain reasonable performance even when only a small number of blocks are received. However, this comes at the cost of reduced performance when most blocks are received (i.e., under good channel conditions), compared to the case where $\epsilon$ is smaller.
    \item When $\epsilon_1<\dots<\epsilon_K$: A strictly increasing sequence of erasure probabilities allows for pre-defined, non-uniform importance across blocks. The choice of the initial value and the rate of increase (i.e., the slope) affects whether the model is biased toward robustness in low-quality channels or enhanced performance in high-quality channels. For example, setting $\epsilon_1=0$  ensures that the first block is always received during training, encouraging the model to concentrate essential information in that block. This improves performance when the first block is reliably received but leads to a sharp drop in quality if it is lost. A steeper slope prompts the encoder to prioritize critical information early (similar to a successive refinement strategy), whereas a gentler slope allows the information to be distributed more evenly, anticipating the reception of subsequent blocks. Our method supports flexible slope design to suit the expected channel behavior.
\end{enumerate}

\subsection{Applications: Semantic network operations that can benefit from the proposed source code}\label{sec:applications}

We target scenarios where limited bandwidth prevents the delivery of all data, resulting in inevitable block losses. Conventional solutions often rely on re-transmission mechanisms to recover the missing data. However, delay-sensitive applications — such as video conferencing— cannot tolerate the latency introduced by retransmissions, which can significantly degrade the quality of service. We discuss three scenarios where the proposed JSCC-based error concealment mechanism can be beneficial:

\begin{itemize}
    \item The first scenario involves settings where a random subset of blocks is dropped due to the bandwidth constraints. In such cases, our approach, by \textit{setting the importance of all blocks to be equal}, enables the decoder to recover the sample from the received subset, leveraging the redundancy and error concealment built into the encoding.
    \item The second scenario addresses network congestion, where intermediate nodes are responsible for deciding which packets to drop. In this setting, our method, by \textit{assigning non-uniform importance to blocks}, allows the network to intelligently discard less important packets when resources are constrained, preserving the quality of reconstruction.
    \item The third scenario integrates the proposed mechanism with physical layer error protection. Specifically, the Modulation and Coding Scheme (MCS) in 5G, which combines modulation order and coding rate, allows networks to control the reliability (i.e., drop probability) of packet delivery. Typically, retransmissions compensate for losses. We propose that, by \textit{setting non-uniform importance levels}, the network can apply stronger protection (lower MCS) to more important blocks and weaker protection (higher MCS) to less critical ones. This intelligent allocation of protection resources ensures timely and reliable delivery of essential content, even under loss or erasure conditions.
\end{itemize}

\section{Experiments}
\label{sec:experiments}

\subsection{Image}

\begin{figure}
    \centering
    \includegraphics[width=1\linewidth]{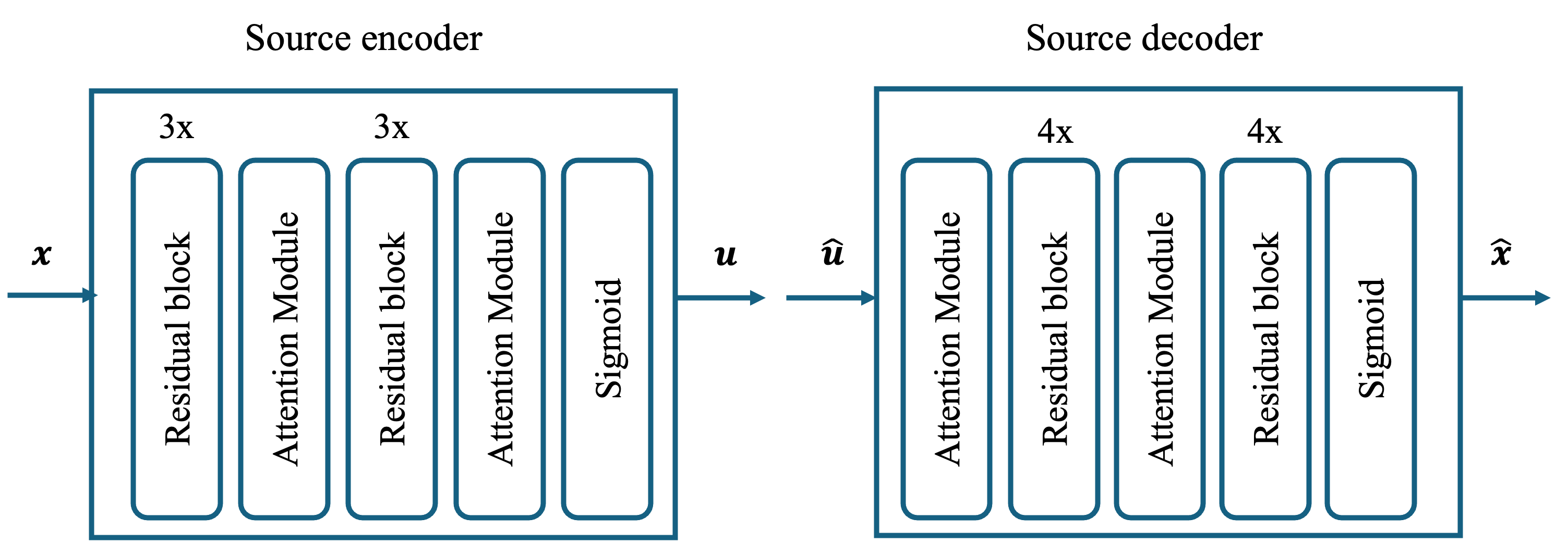}
    \caption{Architecture of the encoder and decoder. 
    \vspace{-0.5cm}}
    \label{fig:ArchitectureImage_simplified}
\end{figure}

The architecture of the model used as the source code is illustrated in Fig.~\ref{fig:ArchitectureImage_simplified}. The models were trained on the CIFAR-10 dataset \cite{krizhevsky09_learn_multip_layer_featur_tiny_images}, using the mean squared error between the original and reconstructed samples as the loss function. We employed the Adam optimizer with a learning rate of $10^{-4}$, and trained the models for $200$ epochs. The resulting encoded representation has a size of $16\times8\times8$, and we set the JSCC parameters as $K=8$, $\alpha=2$, $\beta=8$, and $\gamma=8$.

\begin{figure}
    \centering
    \includegraphics[width=0.99\linewidth]{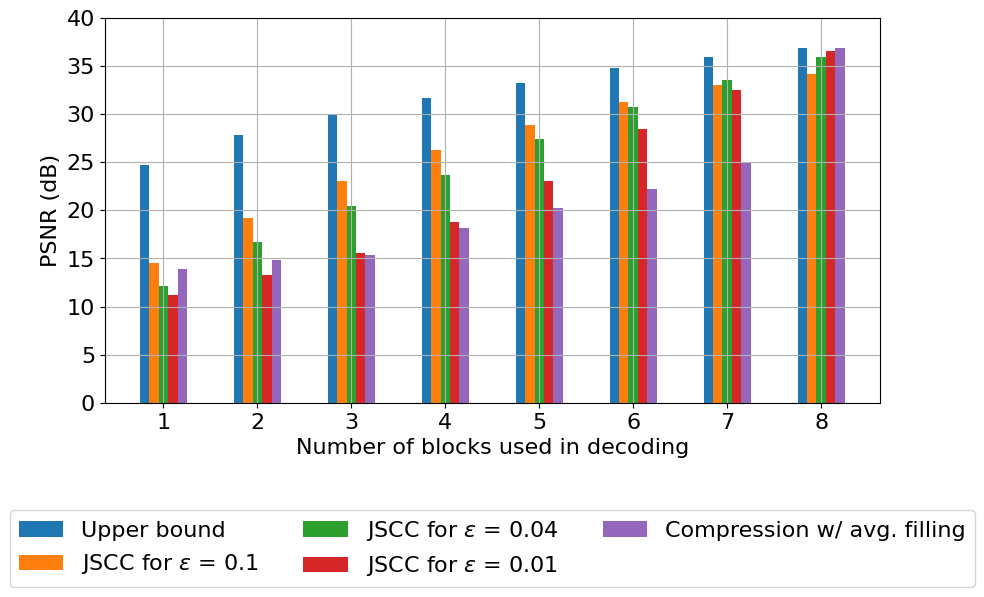}
    \caption{Performance of the proposed method trained under various uniform erasure probabilities and evaluated across different bandwidths.}
    \label{fig:img-application1-subset_decoding}
\end{figure}

We first evaluate the proposed method in a setting where a random subset of $E$ out of $K$ blocks is erased. Training is performed with a uniform erasure probability across all blocks, i.e., $\epsilon_i = \epsilon$ for all $i \in \{1, \dots, K\}$. Fig.~\ref{fig:img-application1-subset_decoding} shows the error concealment performance for different values of training $\epsilon$. The horizontal axis denotes the number of blocks used for decoding (i.e., $K$-$E$), and the vertical axis reports the average PSNR. 
The blue bar provides an upper bound, corresponding to a code with the same architecture trained using only $K-E$ blocks and assuming no erasures during training. Such a scheme would be infeasible since it assumes the erased blocks were known apriori, which is why it serves as a genie-aided upper bound. The purple bar represents a compression scheme where the encoder compresses all samples to $K=8$ blocks without erasure-aware training, thus labelled as ``8-block compression'', and the decoder replaces erased blocks by averaging the received ones. The results reveal a trade-off: larger $\epsilon$ improves robustness when few blocks are received, while smaller $\epsilon$ achieves higher PSNR when all blocks are successfully delivered. Importantly, the performance loss under good channel conditions is minimal for small $\epsilon$, whereas the added redundancy substantially enhances robustness in adverse conditions.

\begin{figure}
    \centering
    \includegraphics[width=0.99\linewidth]{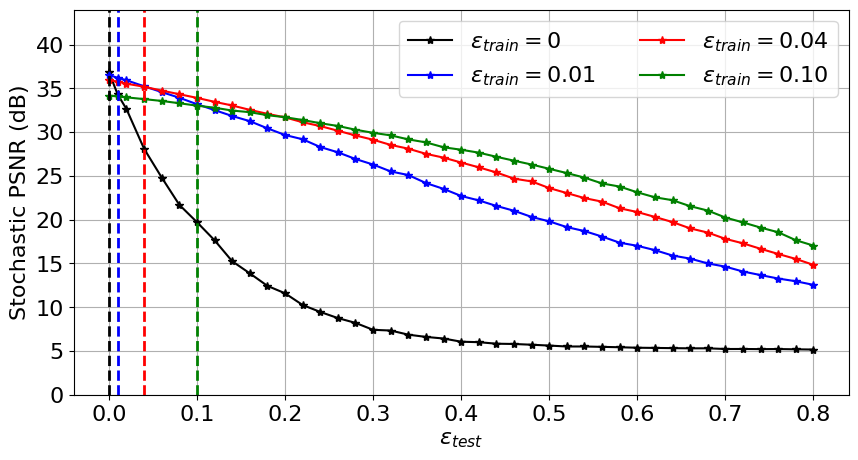}
    \caption{Mismatch analysis showing the average PSNR for codes trained with different $\epsilon_\text{train}$ and tested under varying $\epsilon_\text{test}$. Vertical lines indicate matched training and test conditions.}
    \label{fig:img-mismatch-analysis}
\end{figure}

Next, we perform a mismatch analysis in Fig.~\ref{fig:img-mismatch-analysis}, evaluating performance of the codes trained under uniform erasure probabilities $\epsilon_i = \epsilon_\text{train}$, but tested with a different uniform erasure rate $\epsilon_i = \epsilon_\text{test}$ for all $i \in \{1, \dots, K\}$. The horizontal axis in the figure represents $\epsilon_\text{test}$, while the vertical axis shows the average PSNR over the test dataset. Each curve corresponds to a different value of $\epsilon_\text{train}$. For reference, vertical dashed lines indicate the points where $\epsilon_\text{train} = \epsilon_\text{test}$, i.e., no mismatch between training and testing. As expected, the code trained without error resilience ($\epsilon_\text{train} = 0$) exhibits sharp performance degradation as $\epsilon_\text{test}$ increases. In contrast, codes trained with $\epsilon_\text{train} > 0$ degrade more gracefully under increasing test-time erasure. Notably, all curves show a steeper decline in performance once $\epsilon_\text{test} > \epsilon_\text{train}$. Furthermore, codes trained with higher $\epsilon_\text{train}$ perform better at larger test erasure rates, while they incur a slight performance loss when the test conditions are better (i.e., lower $\epsilon_\text{test}$).

\begin{figure}
    \centering
    \includegraphics[width=1\linewidth]{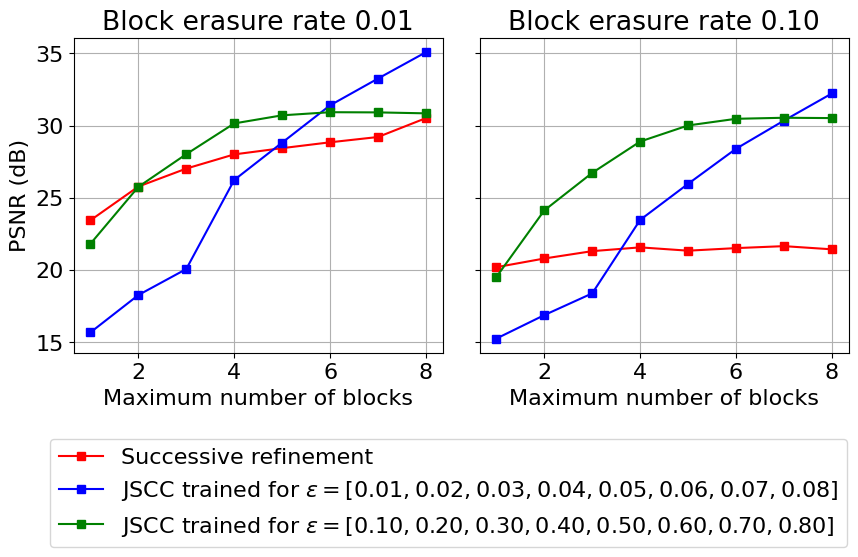}
    \caption{{Performance of the proposed method under increasing block erasure probabilities during training, compared with successive refinement, with a fixed maximum number of transmitted blocks. Each block is independently erased with probability 1\% (left) and 10\% (right).}}
    \label{fig:intelligent_congestion_control}
\end{figure}

{
In a second application, due to resource constraints, only a subset of blocks per sample can be transmitted. During training, we use nonuniform erasure probabilities so that the network can prioritize the most important blocks under limited resources. At inference time, when packet dropping is required, the network removes the last blocks. In addition, due to stochastic effects, each transmitted block is still subject to a nonzero independent erasure probability. As a baseline, we consider successive refinement, where the number of encoder–decoder pairs equals the number of refinements. The first code compresses the input into one block, and each subsequent code adds one block given the previous blocks. This greedy strategy achieves optimal performance under the most constrained bandwidth (i.e., when only one block is transmitted), but becomes suboptimal as bandwidth increases. For a fair comparison, all models are trained using the same architectures and training procedures. Fig.~\ref{fig:intelligent_congestion_control} shows the performance of our method for different training erasure probability vectors. The horizontal axis denotes the maximum number of blocks transmitted per sample, and the vertical axis reports the average PSNR.}

{
The results show that a steeper training erasure profile (green curves) biases the model toward low-quality channels (smaller bandwidths), degrading performance under high-quality conditions. Conversely, a shallower profile (blue curve) favors higher bandwidth performance at the cost of reduced robustness at smaller bandwidths. When the erasure rate is higher (10\%), our method significantly outperforms successive refinement (red curve). This is because successive refinement is highly sensitive to non-ordered block erasures: if an early block is lost, the decoder needs to discard later blocks. Even at a lower erasure rate (1\%), our method trained with a shallower erasure slope significantly outperforms SR when more than half of the blocks are transmitted, as it is non-greedy and anticipates the availability of future blocks. In contrast, the steep-slope variant performs only slightly better than SR, as it also concentrates most information in the first few blocks. Finally, in terms of complexity, the proposed method requires a single encoder–decoder pair that operates with any number of received blocks, whereas successive refinement requires multiple encoder–decoder pairs, one for each refinement layer (eight in this experiment).}

\begin{figure}
    \centering
    \includegraphics[width=0.99\linewidth]{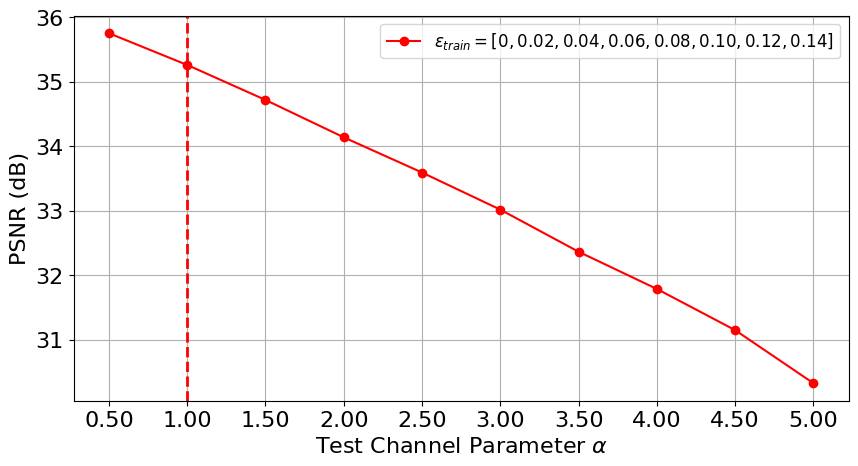}
    \caption{Average PSNR as a function of the test channel severity, where both the training and test erasure probability vectors follow the same pattern thanks to the unequal error protection. vertical line shows the matched case.\vspace{-0.4cm} 
    }
    \label{fig:nonuniform_error_protection_image}
\end{figure}

In the final application, unequal error protection is applied to different blocks of the encoded sample by using a non-uniform erasure probability vector during source code training. To model unequal error protection (e.g., using different MCS levels per block), the actual erasure probability for each block in the test phase is defined as

$$\epsilon_\text{test}=\min(a*\epsilon_\text{train},1),$$
where $a\in\{0.5,1,\dots,5\}$ is a scalar that controls the severity of the test channel. The minimum is applied element-wise, meaning any value greater than $1$ is capped at $1$.  

Fig.~\ref{fig:nonuniform_error_protection_image} shows the resulting average PSNR versus parameter $a$. As observed, unequal error protection leads to a smoother degradation in performance as channel conditions worsen, due to the alignment of the test channel's erasure pattern with the non-uniform training vector.

\subsection{Video}

This section builds upon Deep Video Compression (DVC)~\cite{lu2019dvc}, an end-to-end deep model that optimizes the components of a predictive video compression system {and relies on inferring and transmitting motion and residual vectors per frame, consistent with the standards of H.264/H.265.} 
We expand upon DVC to integrate the compression and channel protection operations, explicitly modeling channel impairments in the motion and residual features during the end-to-end training for video transmission. Our framework is trained on Vimeo-90k~\cite{xue2019video} using the Adam optimizer (initial learning rate $10^{-4}$, and reduction by a factor of $10$ once the loss stabilizes) and a mini-batch size of $50$.
In the proposed system, the motion and feature tensors are each divided into $8$ blocks of size $\frac{C}{8} \times H \times W$, where $C$ is the number of channels, $H$ is the frame height, and $W$ is the frame width.
These blocks pass through a multi-level block erasure channel abstraction, which can be either {uniform} or {non-uniform}. 
The performance is evaluated on the UVG dataset~\cite{mercat2020uvg}.

\begin{figure}
    \centering
    \includegraphics[width=1\linewidth]{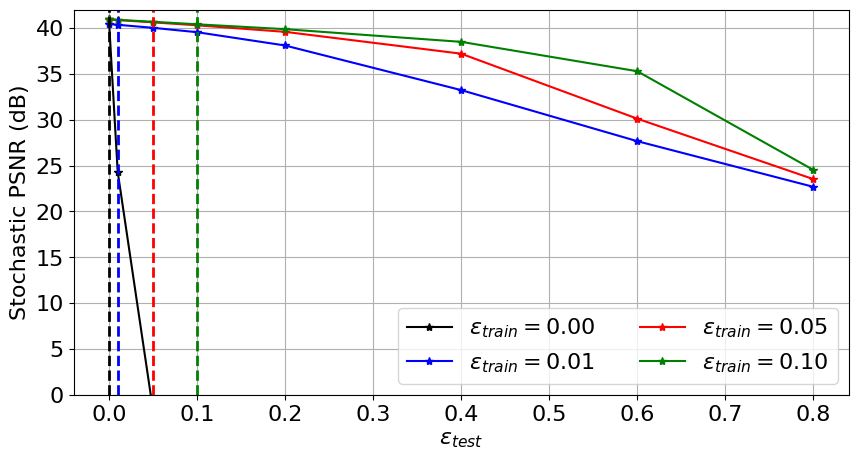}
    \caption{Mismatch analysis in the video domain and for the uniform erasure rate. Vertical lines indicate matched conditions.\vspace{-0.3cm}}
    \label{fig:video_mismatch}
\end{figure}

\begin{figure*}[t]
    \centering
    \begin{subfigure}[b]{0.24\textwidth}
        \includegraphics[width=\textwidth]{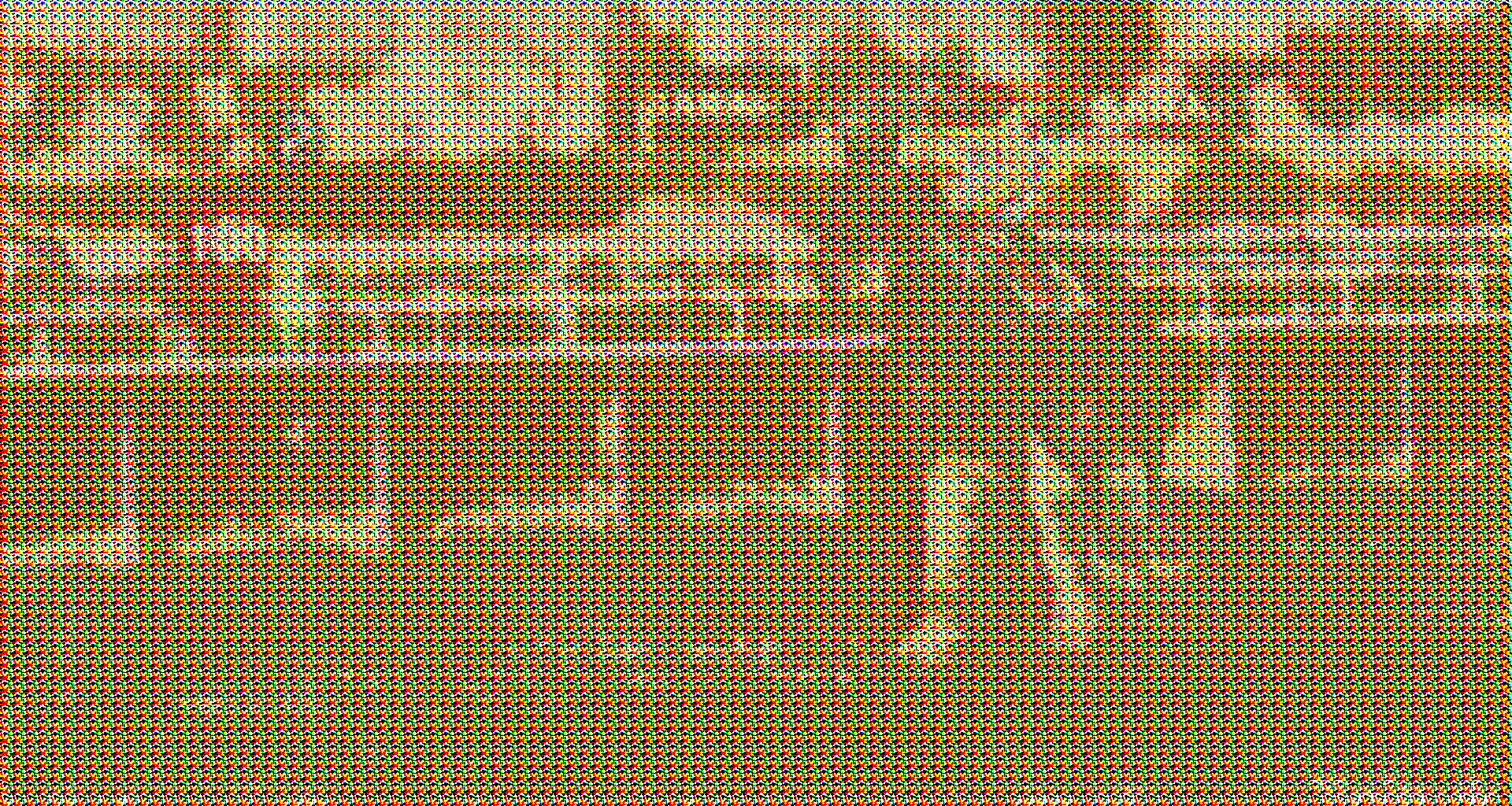}
        \caption{$\epsilon_{\text{train}}=0$, $\epsilon_{\text{test}}=0.1$}
    \end{subfigure}
    \hfill
    \begin{subfigure}[b]{0.24\textwidth}
        \includegraphics[width=\textwidth]{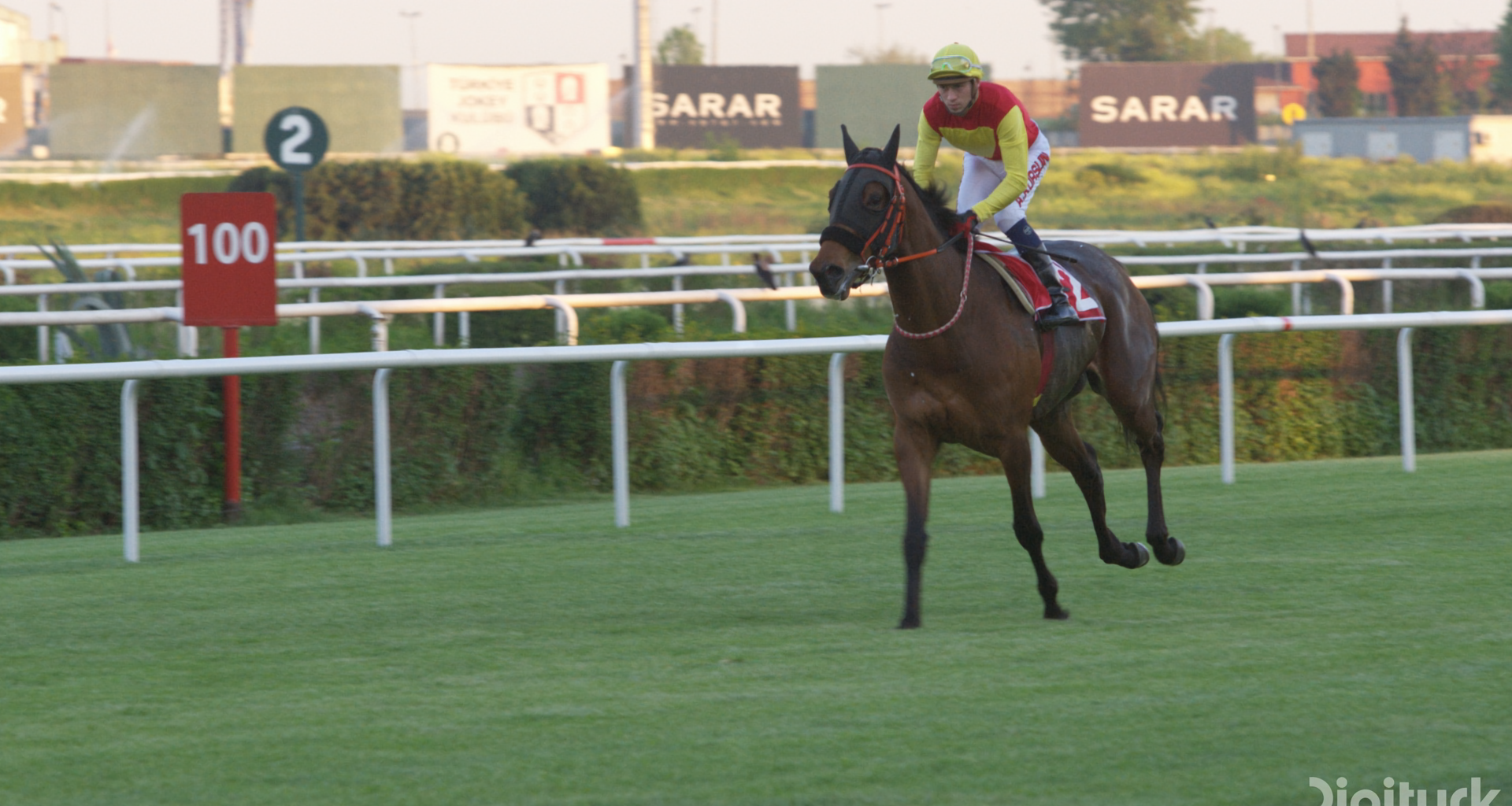}
        \caption{$\epsilon_{\text{train}}=0.1$, $\epsilon_{\text{test}}=0.1$}
    \end{subfigure}
    \hfill
    \begin{subfigure}[b]{0.24\textwidth}
        \includegraphics[width=\textwidth]{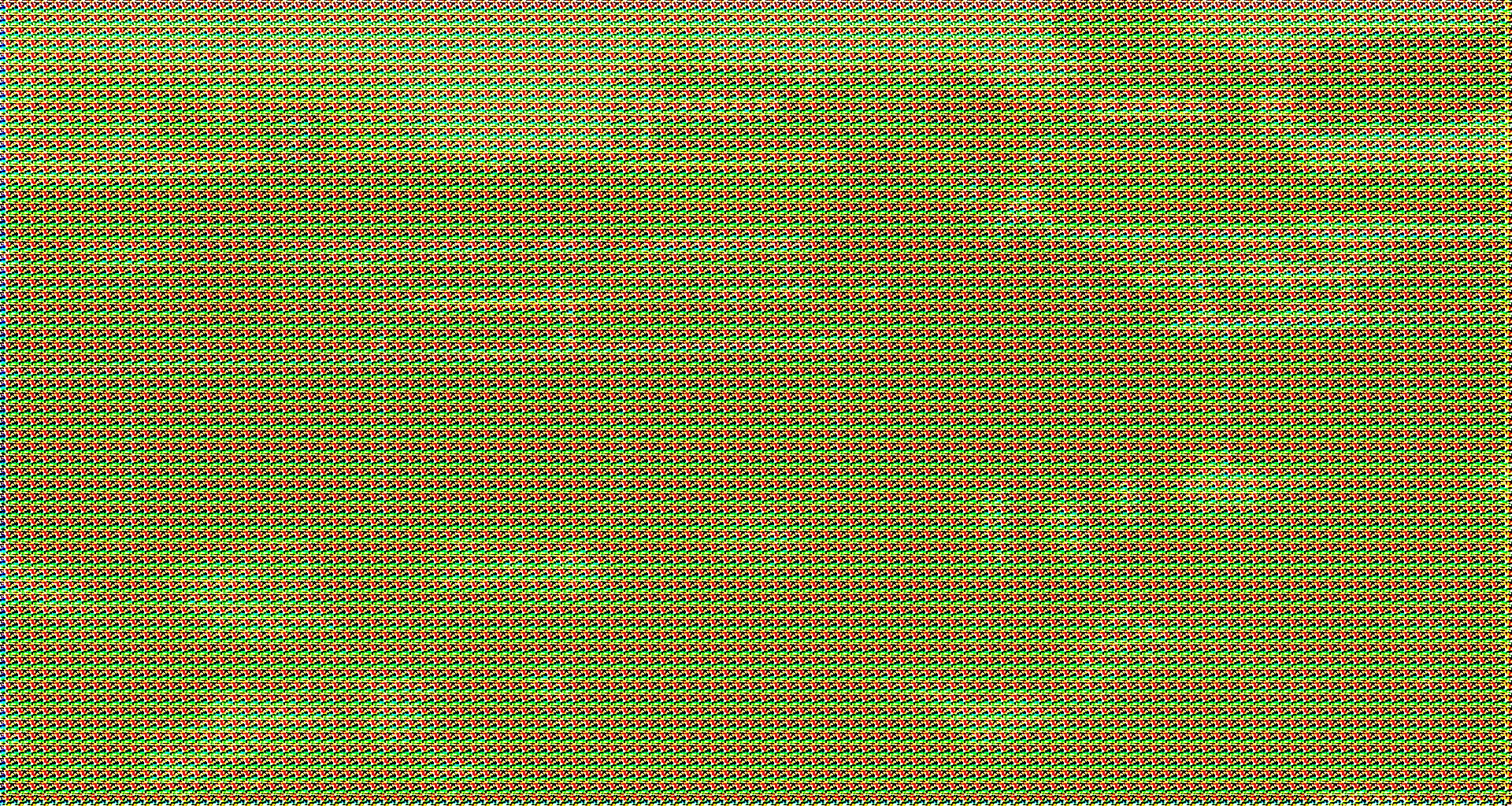}
        \caption{$\epsilon_{\text{train}}=0$, $\epsilon_{\text{test}}=0.4$}
    \end{subfigure}
    \hfill
    \begin{subfigure}[b]{0.24\textwidth}
        \includegraphics[width=\textwidth]{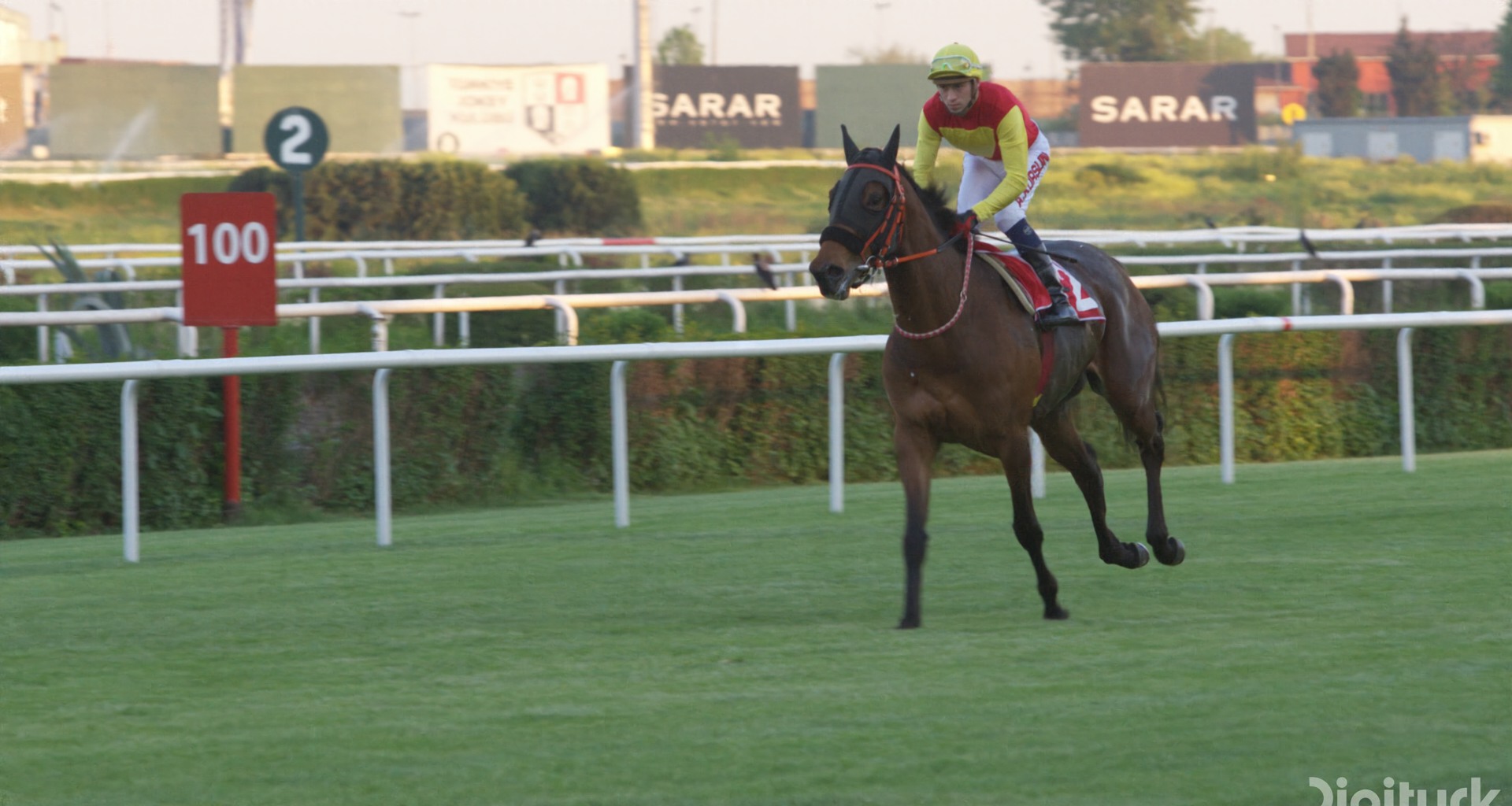}
        \caption{$\epsilon_{\text{train}}=0.1$, $\epsilon_{\text{test}}=0.4$}
    \end{subfigure}

    \vspace{0.3cm} 

    \begin{subfigure}[b]{0.24\textwidth}
        \includegraphics[width=\textwidth]{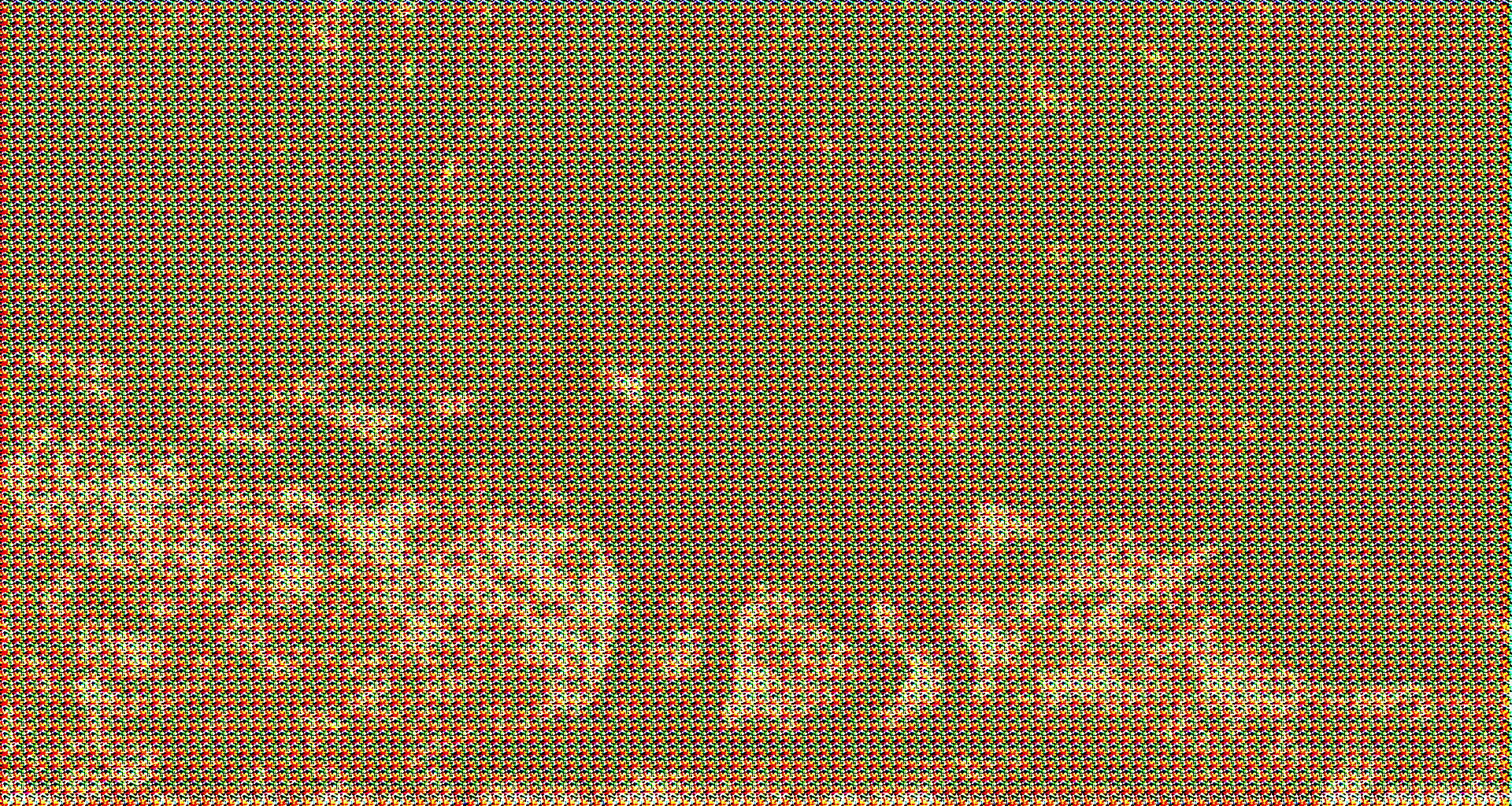}
        \caption{$\epsilon_{\text{train}}=0$, $\epsilon_{\text{test}}=0.1$}
    \end{subfigure}
    \hfill
    \begin{subfigure}[b]{0.24\textwidth}
        \includegraphics[width=\textwidth]{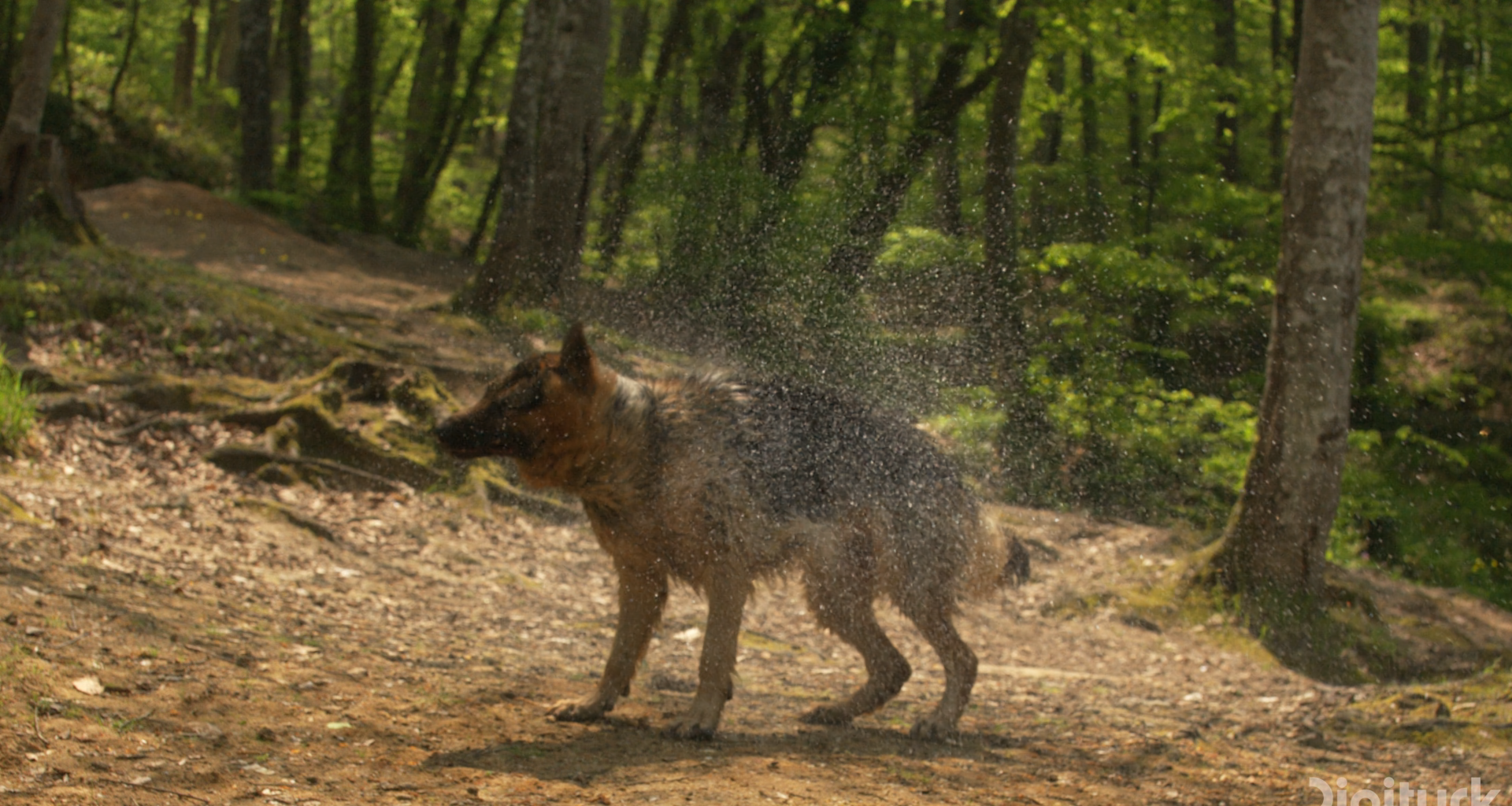}
        \caption{$\epsilon_{\text{train}}=0.1$, $\epsilon_{\text{test}}=0.1$}
    \end{subfigure}
    \hfill
    \begin{subfigure}[b]{0.24\textwidth}
        \includegraphics[width=\textwidth]{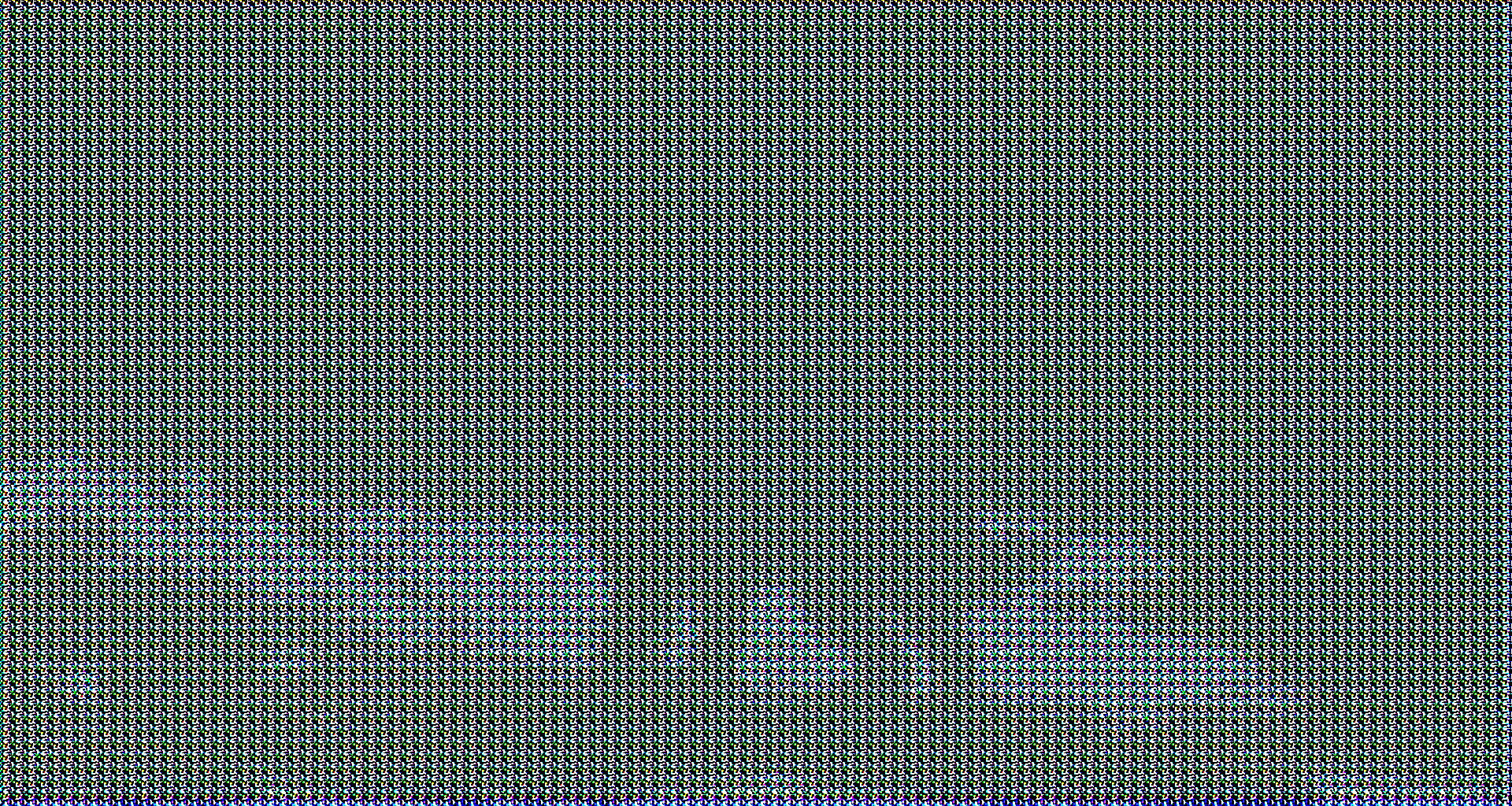}
        \caption{$\epsilon_{\text{train}}=0$, $\epsilon_{\text{test}}=0.4$}
    \end{subfigure}
    \hfill
    \begin{subfigure}[b]{0.24\textwidth}
        \includegraphics[width=\textwidth]{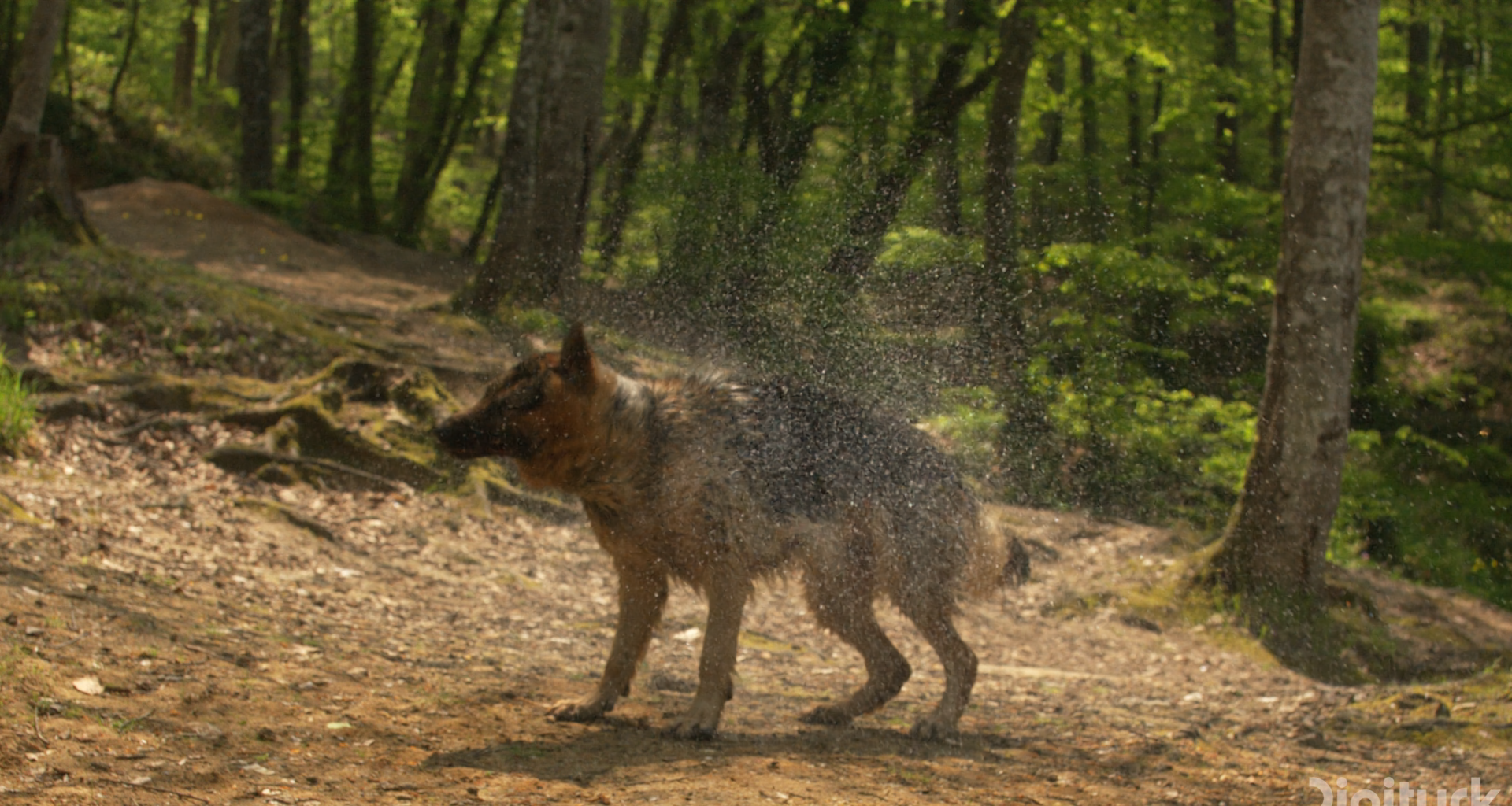}
        \caption{$\epsilon_{\text{train}}=0.1$, $\epsilon_{\text{test}}=0.4$}
    \end{subfigure}

    \caption{Visual comparison of reconstruction results for the base DVC system and one trained for uniform block erasure rate, at two test erasure levels on two video sequences.
    }
    \label{fig:erasure_visual_comparison}
\end{figure*}

We begin by performing a mismatch analysis, analogous to that for the image domain. Fig.~\ref{fig:video_mismatch} illustrates the average frame PSNR as a function of the test erasure rate $\epsilon_{\text{test}}$, for DVC-based JSCC solutions trained with different uniform erasure rates, including the baseline DVC. The baseline DVC model is not designed for erasure resilience, and thus it exhibits a pronounced performance degradation even at very low test erasure rates, reaching negative PSNR values at higher erasure levels. In contrast, models explicitly trained to tolerate block erasures demonstrate remarkably superior performance, with their PSNR gradually decreasing as $\epsilon_{\text{test}}$ increases. Notably, the performance loss under good channel conditions, incurred by embedding redundancies, is incremental in this experiment. This leads to the model trained with $\epsilon_{\text{train}} = 0.10$ achieving the greatest robustness, exhibiting minimal performance degradation up to $\epsilon_{\text{test}} \approx 0.3$ and a smooth decline thereafter. For the visual comparison in Fig.~\ref{fig:erasure_visual_comparison}, we compare the baseline system which has $\epsilon_{\text{train}}=0$ with the system trained under uniform erasure rate $\epsilon_{\text{train}}=0.1$. The results clearly demonstrate the advantage of the proposed system across different levels of erasure.

\begin{figure}
    \centering
    \includegraphics[width=1\linewidth]{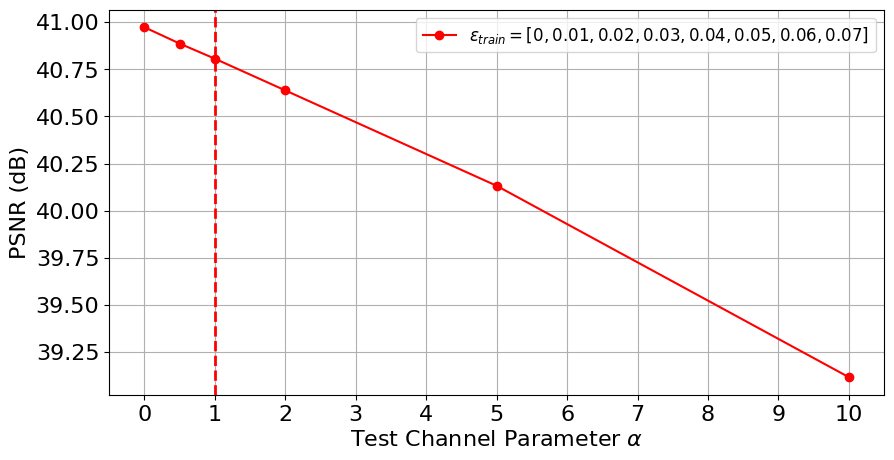}
   \caption{Reconstruction performance under non-uniform block erasure rate with varying $\epsilon_{\text{train}}$ and scaling parameter $\epsilon_{\text{test}}$. 
   }
    \label{fig:video_non_uniform}
\end{figure}

In the non-uniform erasure rate scenario, both the motion and residual blocks of the video compression system were trained using progressively increasing erasure probabilities, specifically
$\epsilon_{\text{train}} = [0, 0.01, 0.02, 0.03, 0.04, 0.05, 0.06, 0.07]$. 
The trained models were then evaluated against test erasure rates defined as $\epsilon_{\text{test}} = \min(a* \epsilon_{\text{train}},1)$, where 
$a \in \{0, 0.5, 1, 2, 5, 10\}$.
Fig.~\ref{fig:video_non_uniform} illustrates the average PSNR as a function of the scaling parameter $a$. As depicted in the plot, the average PSNR maintains near-peak values for small $a$, indicating accurate video frame reconstruction when test erasures are within or slightly exceed the training range. While PSNR gradually decreases with increasing $a$, even at $a = 10$, the PSNR remains above $39$~dB. This trend highlights the effectiveness of progressive, erasure rates and unequal error protection in ensuring high-fidelity reconstruction in the presence of severe, and previously unseen, channel conditions.

\begin{figure}
    \centering
    \includegraphics[width=1\linewidth]{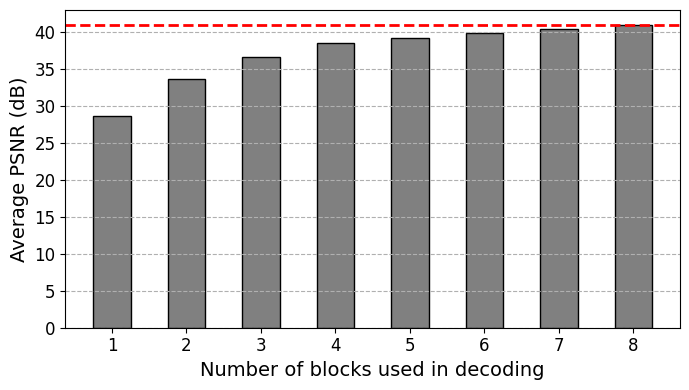}
   \caption{Average PSNR versus number of decoding blocks under non-uniform erasure rate vector for training.}
    \label{fig:nu_decoding}
\end{figure}

\begin{figure}[t]
    \centering
    \begin{subfigure}[b]{0.45\textwidth}
        \includegraphics[width=\textwidth]{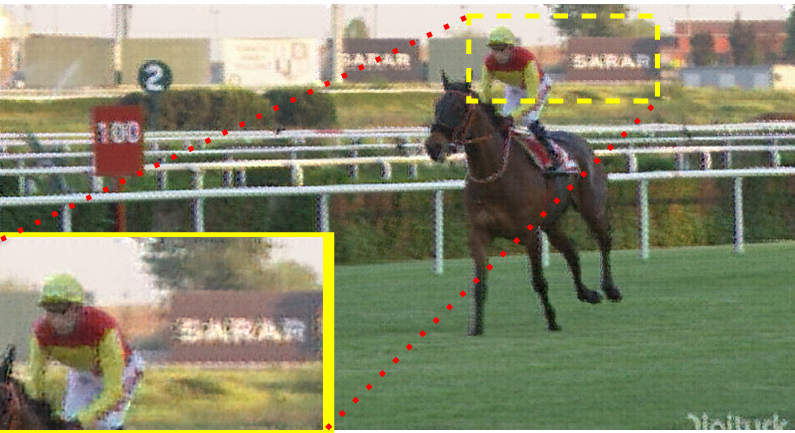}
        \caption{Reconstruction using $1$ block (PSNR= $28.11$ dB)}
    \end{subfigure}
    \hfill
    \begin{subfigure}[b]{0.45\textwidth}
        \includegraphics[width=\textwidth]{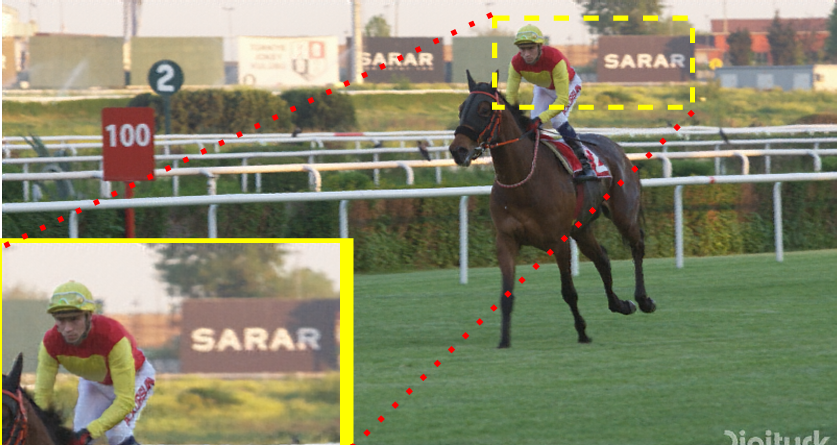}
        \caption{Reconstruction using $2$ blocks (PSNR= $35.46$ dB)}
    \end{subfigure}
    \hfill
    \begin{subfigure}[b]{0.45\textwidth}
        \includegraphics[width=\textwidth]{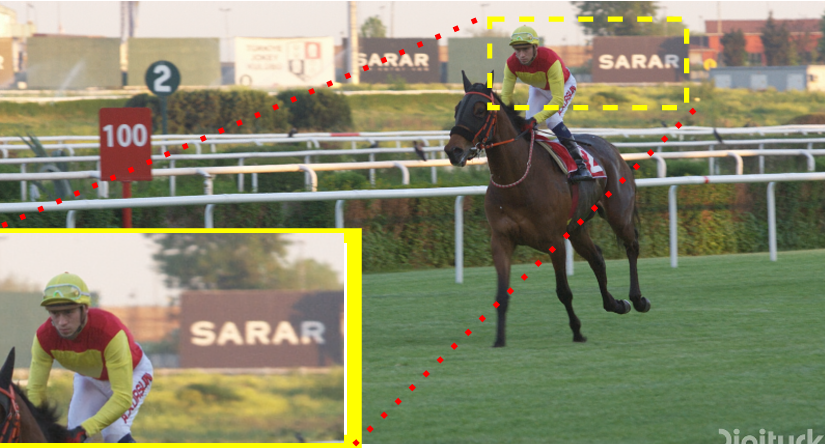}
        \caption{Reconstruction using $4$ blocks (PSNR= $39.55$ dB)}
    \end{subfigure}

    \caption{Visual comparison of the performance in the non-uniform erasure scenario when only partial blocks are received. }
    \label{fig:nu_block_vector}
\end{figure}

Fig.~\ref{fig:nu_decoding} illustrates that under non-uniform erasure rates, PSNR significantly increases with the number of decoding blocks, rising from approximately $29$~dB (single block) to nearly $41$~dB (all eight blocks), with the red dashed line indicating peak quality. This reflects a hierarchical information structure: early blocks encode critical coarse content and are better protected, yielding steep initial PSNR gains, while later blocks contribute finer details with diminishing returns. Fig.~\ref{fig:nu_block_vector} visually supports this, highlighting the system's ability to maintain high-quality outputs despite partial block reception and severe erasure conditions.

\section{Conclusion}

This paper presents a solution capable of supporting next-generation applications that demand low-latency image/video transmission in 6G networks and future wireless systems. By enabling importance-aware and adaptive data processing within the network, our method eliminates the need for explicit semantic sharing or feedback to the source encoder. This approach is particularly beneficial for critical applications such as video conferencing and robotic remote control.




\bibliographystyle{IEEEtran} 
\bibliography{references}

\end{document}